\documentclass[aps,10pt,pra,notitlepage,twocolumn,showpacs,superscriptaddress]{revtex4-1}%
\usepackage{amsfonts}
\usepackage{amsmath}
\usepackage{amssymb}
\usepackage{graphicx}%

\providecommand{\U}[1]{\protect\rule{.1in}{.1in}}
\providecommand{\U}[1]{\protect\rule{.1in}{.1in}}

\begin{document}
\preprint{ }
\title{The roton-assisted chiral $p$-wave superfluid in a quasi-two-dimensional
dipolar Bose-Fermi quantum gas mixture}
\author{Ben Kain}
\affiliation{Department of Physics, College of the Holy Cross, Worcester, Massachusetts 01610, USA}
\author{Hong Y. Ling}
\affiliation{Department of Physics and Astronomy, Rowan University, Glassboro, New
Jersey 08028, USA}
\affiliation{Kavli Institute for Theoretical Physics, University of California, Santa Barbara, California 93106, USA}

\begin{abstract}
\noindent The chiral p-wave ($p_{x}\pm ip_{y}$) superfluid has attracted significant
attention in recent years, mainly because its vortex core supports a Majorana
fermion which, due to its non-Abelian statistics, can be explored for
implementing topological quantum computation. Mixing dipolar bosons with
fermions in quasi-two-dimensional (2D) space offers the opportunity to use the
roton minimum as a tool for engineering the phonon-induced attractive
interaction between fermions. \ We study, within the Hartree-Fock-Bogoliubov
approach, the $p$-wave superfluid pairings in a quasi-2D dipolar Bose-Fermi
mixture. \ We show that enhancing the induced interaction by lowering the
roton minimum can affect the stability property of the mixture as well as the
effective mass of the fermions in an important way. We also show that one can
tune the system to operate in stable regions where chiral $p$-wave superfluid
pairings can be resonantly enhanced by lowering the energy cost of the phonons
near the roton minimum.

\end{abstract}

\pacs{03.75.Ss, 74.20.Rp, 67.30.H-, 05.30.-d}
\maketitle

\section{Introduction}

The chiral $p$-wave ($p_{x}\pm$ $ip_{y}$)\ phase in both superconductors
\cite{reed00PhysRevB.61.10267,ivanov01PhysRevLett.86.268,stern04PhysRevB.70.205338,stone06PhysRevB.73.014505,fu08PhysRevLett.100.096407,sau10PhysRevLett.104.040502}
and ultracold atomic systems
\cite{gurarie05PhysRevLett.94.230403,cheng05PhysRevLett.95.070404,cooper09PhysRevLett.103.155302,tewari07PhysRevLett.98.010506,zhang08PhysRevLett.101.160401,nishida09AnnPhys.324.897,levinsen11PhysRevA.84.013603}
has attracted significant attention in recent years, mainly because its vortex
core supports a unique bound quasiparticle (non-Abelian) excitation with zero
energy known as a Majorana fermion. A two-dimensional (2D) topological system
with well isolated vortices can be thought of as a quantum computing device
where a unitary transformation (braiding vortices \cite{kitaev06AnnPhys.321.2}%
), the basic operation of any quantum computer, is performed in a degenerate
space \cite{ivanov01PhysRevLett.86.268,nayak96NuclPhysB.479.529} topologically
protected from other nontopological excited states by an energy gap
\cite{kitaev06AnnPhys.321.2}. In chiral $p$-wave superfluid systems, this
energy gap is proportional to the $p$-wave superfluid gap parameter.
\ Increasing the $p$-wave superfluid pairing, then, can increase the energy
gap, which, in turn, makes the topologically protected computation less
susceptible to finite temperature effects. \ However, for a wave with finite
angular momentum, there is a centrifugal barrier in its scattering potential
which prevents the wave from penetrating the interior and scattering
effectively from the two-body potential. \ Thus, for typical ground state
atoms, the $p$-wave superfluid pairing is highly suppressed.

The advent of cold-atom systems opened up the possibility of\ using
electromagnetic fields to tune and design two-body interactions capable of
enhancing chiral $p$-wave superfluid pairings
\cite{gurarie05PhysRevLett.94.230403,cheng05PhysRevLett.95.070404,cooper09PhysRevLett.103.155302,tewari07PhysRevLett.98.010506,zhang08PhysRevLett.101.160401,nishida09AnnPhys.324.897,levinsen11PhysRevA.84.013603}%
. The same goal may also be achieved in a Bose-Fermi mixture since bosons can
induce and hence modify the Fermi-Fermi interaction in a Fermi gas
\cite{efremov02PhysRevB.65.134519,wang05PhysRevA.72.051604}. \ \ In a dipolar
quantum gas \cite{santos00PhysRevLett.85.1791,yi00PhysRevA.61.041604}, the
dipole-dipole interaction represents a control knob inaccessible to nondipolar
bosons. Thus, mixing dipolar bosons with fermions opens up new possibilities
\cite{dutta10PhysRevA.81.063608,kain11PhysRevA.83.061603,kain12PhysRevA.85.013631}%
. \ Experimentally, great progress has been made in recent years in achiving
dipolar quantum gases consisting of either heteronuclear molecules with
electric dipoles \cite{ni08Science.322.5899,ospelkaus09FaradayDiscuss142.351}
or atoms with magnetic dipoles
\cite{vengalattore08PhysRevLett.100.170403,stuhler05PhysRevLett.95.150406,lu11PhysRevLett.107.190401,aikawa12PhysRevLett.108.210401}%
. \ In this article, we consider a quasi-two-dimensional (2D) cold-atom mixture
of nondipolar fermions of mass $m_{F}$ in a spin-polarized Fermi gas and
bosons of mass $m_{B}$ in a dipolar condensate where all the dipoles are
oriented along the axial ($z$) direction.  The collective excitation of such a system was discovered \cite{santos03PhysRevLett.90.250403} to
exhibit, at finite momentum, a minimum analogous to the ``roton" minimum in superfluid helium \cite{landau47JPhysUSSR.11.91,feynman54PhysRev.94.262}.  Dutta and Lewenstein
\cite{dutta10PhysRevA.81.063608} pointed out recently that the existence of
dipolar bosons in such a model can significantly enhance the unconventional
superfluids in the spin-polarized Fermi gas. \ The underlying physical
mechanism \cite{dutta10PhysRevA.81.063608} is the ability to enhance the
induced Fermi-Fermi attraction by lowering the energy cost of the phonons of
the dipolar condensate near the roton minimum
\cite{santos03PhysRevLett.90.250403} (see the main text for detail). \ This
raises the questions of how\ the enhanced induced interaction, usually ignored
in the stability analysis of the Bose-Fermi mixture, affects the stability
property of the mixture, and how this same enhanced interaction affects the
effective mass of the fermions near the Fermi energy and hence the $p$-wave
superfluid critical temperature. \ In what follows, we apply the
Hartree-Fock-Bogoliubov mean-field theory to gain insight into these questions
and identify the parameter space in which the mixture is stable against
density fluctuations and the $p$-wave superfluid pairing can be resonantly enhanced.

\section{Quasi-2D dipolar condensate and its excitation spectrum}

To begin with, we divide the interactions (in momentum space) into a
short-range part, $U_{BF}^{3D}=4\pi\hbar^{2}a_{BF}/m_{BF}$ and $U_{BB}^{3D}$
$=4\pi\hbar^{2}a_{BB}/m_{B}$, characterized with s-wave scattering lengths
$a_{BF}$, between a boson and a fermion, and $a_{BB}$, between two bosons,
where $m_{BF}=2m_{B}m_{F}/\left(  m_{B}+m_{F}\right)  $, and a long range
part, $U_{DD}^{3D}\left(  \mathbf{k}\right)  =8\pi d^{2}P_{2}\left(
\cos\theta_{\mathbf{k}}\right)  $, due to the dipole-dipole interaction, where
$d$ is the induced dipole moment, $P_{2}\left(  x\right)  $ the second-order
Legendre polynomial, and $\theta_{\mathbf{k}}$ the polar angle of wavevector
$\mathbf{k}$. \ Thus, $U_{BB}^{3D}\left(  \mathbf{k}\right)  =U_{BB}^{3D}$
$+U_{DD}^{3D}\left(  \mathbf{k}\right)  $ is the total Bose-Bose interaction.
\ The particles in the mixture are confined, by sufficiently high harmonic
trap potentials along the axial dimension, to a quasi-2D plane where the axial
degree of freedom is \textquotedblleft frozen\textquotedblright\ to the
zero-point oscillation and can thus be factored out.\ Integrating out the
axial degree of freedom via the relations $U_{BF}=\int n_{B}\left(
k_{z}\right)  U_{BF}^{3D}n_{F}\left(  k_{z}\right)  dk_{z}$ and $U_{BB}\left(
\mathbf{k}_{\perp}\right)  =\int n_{B}\left(  k_{z}\right)  U_{BB}^{3D}\left(
\mathbf{k}\right)  n_{B}\left(  k_{z}\right)  dk_{z}$, where $n_{B,F}\left(
k_{z}\right)  =\exp\left(  -d_{B,F}^{2}k_{z}^{2}/4\right)  /\sqrt{2\pi}$, with
$d_{B,F}$ the axial oscillator lengths, we arrive at a qasi-2D dipolar
Bose-Fermi mixture characterized by the effective 2D interactions
$U_{BF}=U_{BF}^{3D}/\sqrt{\pi\left(  d_{F}^{2}+d_{B}^{2}\right)  },$
$U_{BB}=U_{BB}^{3D}/\sqrt{2\pi}d_{B}$, and
\begin{equation}
U_{BB}\left(  k_{\perp}\right)  =U_{BB}\left[  1+2\varepsilon_{dd}%
-3\varepsilon_{dd}\mathcal{G}\left(  d_{B}k_{\perp}\right)  \right]  ,
\label{UBB(k)}%
\end{equation}
where $\varepsilon_{dd}=4\pi d^{2}/\left(  3U_{BB}^{3D}\right)  $ measures the
dipolar interaction relative to the s-wave contact interaction, and
\begin{equation}
\mathcal{G}\left(  x\right)  =\sqrt{\frac{\pi}{2}}x\exp\left(  \frac{x^{2}}%
{2}\right)  \text{Erfc}\left(  \frac{x}{\sqrt{2}}\right)  \,, \label{g(x)}%
\end{equation}
is a well-behaved function \cite{pedri05PhysRevLett.95.200404} which increases
monotonically from 0 to $1$, reaching $50\%$ of its full value when
$x\approx0.61$, and Erfc$\left(  x\right)  $ is the complementary error
function. \ In arriving at these effective interactions, we have assumed that
the axial oscillator lengths are sufficiently large compared with the relevant
scattering lengths so that atoms experience essentially 3D collisions and
there is no need to consider the confinement induced resonances
\cite{olshanii98PhysRevLett.81.938,petrov00PhysRevLett.84.2551,bergeman03PhysRevLett.91.163201}%
. \ Using the Bogoliubov perturbative ansatz and diagonalization procedure
\cite{bijlsma00PhysRevA.61.053601}, we can describe the dipolar Bose gas near
zero temperature ($T\approx0$) as a uniform dipolar condensate of number
density $n_{B}$ plus a collection of phonon modes that obey the dispersion
spectrum
\begin{equation}
E_{k,B}=\hbar kv_{B}\sqrt{1+2\varepsilon_{dd}+\chi_{B}^{2}d_{B}^{2}%
k^{2}-3\varepsilon_{dd}\mathcal{G}\left(  d_{B}k\right)  },
\label{phonon spectrum}%
\end{equation}
where $v_{B}=\sqrt{n_{B}U_{BB}/m_{B}}$ is the phonon speed and $\chi_{B}%
=\xi_{B}/d_{B}$ measures the healing length, $\xi_{B}=\hbar/\sqrt{4m_{B}%
n_{B}U_{BB}}$, in terms of $d_{B}$. \ (From now on, $\mathbf{k}$ stands for
$\mathbf{k}_{\perp}$, the in-plane wavevector, for notational convenience.)

In a dipolar condensate, there is a competition between the side-to-side
arrangement of two dipoles where the dipole-dipole interaction is the most
repulsive and the head-to-tail configuration where the dipole-dipole
interaction is the most attractive. In the 3D (or large $d_{B}$) limit, the
head-to-tail configuration and hence the attractive interaction can be
significant, and $U_{BB}\left(  k\right)  $ in Eq. (\ref{UBB(k)}) reduces to
$U_{BB}\left(  1-\varepsilon_{dd}\right)  $. In this limit, when
$\varepsilon_{dd}$ $>$ $1$, $E_{k,B}$ in Eq. (\ref{phonon spectrum}) acquires
an imaginary frequency in the long wavelength limit, and the condensate is
unstable against collapse. \ In the 2D (or small $d_{B}$) limit, the
head-to-tail arrangement is suppressed, and $U_{BB}\left(  k\right)  $ in Eq.
(\ref{UBB(k)}) asymptotes to the effective 2D contact interaction
$U_{BB}\left(  0\right)  =U_{BB}\left(  1+2\varepsilon_{dd}\right)  $, which
is $1+2\varepsilon_{dd}$ times more repulsive than if the dipole interaction
were absent. In the quasi-2D case, which is our focus here, the head-to tail
configuration is not completely suppressed. \ This is the origin of the
attractive part in Eq. (\ref{UBB(k)}), which depends on $\mathcal{G}\left(
x\right)  $ defined in Eq. (\ref{g(x)}). \ As a result, in quasi-2D space,
$\varepsilon_{dd}$ $>$ $1$ does not automatically mean condensate collapse as
in the 3D limit. \ Instead, $\varepsilon_{dd}$ $>$ $1$ signals that a roton
minimum \cite{santos03PhysRevLett.90.250403} at finite $k$ can occur for
sufficiently small $\chi_{B}$ (or equivalently sufficiently high boson
density) as illustrated in Fig. \ref{Fig:1}(a), and only when the roton
minimum crosses zero, becoming imaginary, does the condensate become unstable
(against the formation of a density wave). \ 

\section{Stability of the quasi-2D dipolar Bose-Fermi mixture}

Integrating out the phonon degrees of freedom, we obtain a total effective
Hamiltonian, $\hat{H}^{E}=H_{B}^{0}+\hat{H}_{F}^{E}$, for a quasi-2D mixture
with an effective area $A$ and fixed particle number densities, $n_{B}$ and
$n_{F}$, where $H_{B}^{0}=AU_{BB}\left(  0\right)  n_{B}^{2}/2$ accounts for
the dipolar condensate energy, and
\begin{align}
\hat{H}_{F}^{E}  &  =\sum_{\mathbf{k}}\left[  \epsilon_{\mathbf{k}}\left(
=\hbar^{2}k^{2}/2m_{F}\right)  +U_{BF}n_{B}\right]  \hat{a}_{\mathbf{k}}%
^{\dag}\hat{a}_{\mathbf{k}}\nonumber\\
&  +\frac{1}{2A}\sum_{\mathbf{k},\mathbf{k}^{\prime},\mathbf{q}}U\left(
q\right)  \hat{a}_{\mathbf{k}+\mathbf{q}}^{\dag}\hat{a}_{\mathbf{k}^{\prime
}-\mathbf{q}}^{\dag}\hat{a}_{\mathbf{k}^{\prime}}\hat{a}_{\mathbf{k}}\text{,}
\label{HFE}%
\end{align}
describes an effective Fermi gas in which fermions interact via the two-body
interaction potential (in the static limit
\cite{bijlsma00PhysRevA.61.053601,efremov02PhysRevB.65.134519}), $U\left(
k\right)  =-\left(  U_{BF}^{2}/U_{BB}\right)  \bar{U}\left(  k\right)  $,
where
\begin{equation}
\bar{U}\left(  k\right)  =\frac{1}{1+2\varepsilon_{dd}+\chi_{B}^{2}d_{B}%
^{2}k^{2}-3\varepsilon_{dd}\mathcal{G}\left(  d_{B}k\right)  }.
\label{phonon-induced interaction}%
\end{equation}
Note that Fermi statistics prohibits two fermions in a spin-polarized Fermi
gas to participate in s-wave scattering. \ In our model, two fermions of the
same spin can interact indirectly by exchanging virtual phonons with the help
of the intermediate states containing one phonon without violating the Pauli
exclusion principle. Equation (\ref{phonon-induced interaction}) describes
precisely this phonon-mediated Fermi-Fermi interaction.

The Fermi pairings can benefit from this induced interaction, which is now
tunable by varying the roton minimum, provided that fermions and bosons can
form stable mixtures near the roton minimum. \ Thus, we first address the
stability issue using the energy density, $\mathcal{E}\left(  n_{B}%
,n_{F}\right)  =(H_{B}^{0}+\left\langle \hat{H}_{F}^{E}\right\rangle )/A$,
where, for simplicity, the average is performed assuming that the fermions are
in the normal gas phase (at $T=0$)\ and thus distribute according to
$\left\langle \hat{a}_{\mathbf{k}}^{\dag}\hat{a}_{\mathbf{k}}\right\rangle
=\Theta\left(  k_{F}^{2}-k^{2}\right)  $. $\ \ $Here, $\Theta\left(  x\right)
$ is the step function and $k_{F}=\sqrt{4\pi n_{F}}$ the (2D) Fermi momentum.
\ A straightforward application of the Hartree-Fock mean-field theory
\cite{fetter71ManyParticleSystemsBook} finds,
\begin{align}
\mathcal{E}\left(  n_{B},n_{F}\right)   &  =\frac{U_{BB}\left(  1+2\varepsilon
_{dd}\right)  }{2}n_{B}^{2}+U_{BF}n_{B}n_{F}+\frac{\pi\hbar^{2}}{m_{F}}%
n_{F}^{2}\nonumber\\
&  -\frac{\pi\hbar^{2}}{m_{F}}\frac{\lambda}{1+2\varepsilon_{dd}}n_{F}%
^{2}+\mathcal{E}_{exc}\left(  n_{B},n_{F}\right)  ,\label{energy densities}%
\end{align}
where the first line, which will be denoted as $\mathcal{E}_{0}\left(
n_{B},n_{F}\right)  $, represents the energy of the mixture when the induced
interaction is ignored, and the terms in the second line represent,
respectively, the direct (Hartree) and the exchange (Fock) energy of the
induced interaction Hamiltonian in Eq. (\ref{HFE}). \ It can be shown that the
latter, which is defined as $\mathcal{E}_{exc}=-\sum_{\mathbf{k}%
,\mathbf{k}^{\prime}}U\left(  \left\vert \mathbf{k}-\mathbf{k}^{\prime
}\right\vert \right)  \left\langle \hat{a}_{\mathbf{k}^{\prime}}^{\dag}\hat
{a}_{\mathbf{k}^{\prime}}\right\rangle \left\langle \hat{a}_{\mathbf{k}}%
^{\dag}\hat{a}_{\mathbf{k}}\right\rangle /2A^{2}$, can be reduced to\ the 1D
integral%
\begin{equation}
\mathcal{E}_{exc}\left(  n_{B},n_{F}\right)  =\lambda\frac{\pi\hbar^{2}}%
{m_{F}}\frac{k_{F}^{2}}{\left(  2\pi\right)  ^{3}}\int_{0}^{2k_{F}}\bar
{U}\left(  q\right)  I\left(  q\right)  qdq,\label{exchange energy density}%
\end{equation}
where $I\left(  q\right)  $ is a function of $q$ defined as
\begin{equation}
I\left(  q\right)  =2\sin^{-1}\sqrt{1-\left(  \frac{q}{2k_{F}}\right)  ^{2}%
}-\frac{q}{k_{F}}\sqrt{1-\left(  \frac{q}{2k_{F}}\right)  ^{2}},\label{I(q)}%
\end{equation}
when $q<2k_{F}$ and $I\left(  q\right)  =0$ otherwise. \ \ The quantity
$\lambda$, in both Eqs. (\ref{energy densities}) and
(\ref{exchange energy density}), is a unitless coupling parameter defined as
$\lambda=N\left(  \epsilon_{F}\right)  U_{BF}^{2}/U_{BB}$, where $N\left(
\epsilon_{F}\right)  =m_{F}/2\pi\hbar^{2}$ is the (2D) density of states
independent of the Fermi energy $\epsilon_{F}$. \ The same $\lambda$ also
appears in the gap equation \ (\ref{gap l}) we will use to estimate the
critical temperature.

The uniform mixture is unstable against small changes in densities if any of
the following inequalities, (a) $\partial_{n_{B}}^{2}\mathcal{E}<0$, (b)
$\partial_{n_{F}}^{2}\mathcal{E}<0$, and (c) $\left(  \partial_{n_{B}}%
^{2}\mathcal{E}\right)  \left(  \partial_{n_{F}}^{2}\mathcal{E}\right)
-\left(  \partial_{n_{B},n_{F}}^{2}\mathcal{E}\right)  ^{2}<0$, occur. \ The
analysis below, however, focuses only on the instability towards phase
separation (c) as it always precedes collapse of the condensate (a) and that
of the Fermi gas (b) for the parameters of interest. \ It is well-known, from
stability analysis of the Bose-Fermi mixtures in which the phonon-induced
interaction is insignificant and is thus neglected, that the dimensionality
plays an important role. \ \ This is because the kinetic energy of the Fermi
sea depends on the Fermi surface gemometry, which is strongly influenced by
the dimensionality. The kinetic energy (of the Fermi sea) is $\sim\left(
n_{F}^{3D}\right)  ^{5/3}$ in 3D and $\sim\left(  n_{F}^{1D}\right)  ^{3}$ in
1D. \ Thus, contrary to 3D mixtures, which are stable only when $n_{F}^{3D}$
is below a threshold \cite{viverit00PhysRevA.61.053605}, 1D mixtures are
stable provided that $n_{F}^{1D}$ is above a threshold
\cite{das03PhysRevLett.90.170403}. \ 2D mixtures are distinct in the sense
that the 2D kinetic energy follows the square law, $\sim n_{F}^{2}$. A simple
analysis of $\mathcal{E}_{0}\left(  n_{B},n_{F}\right)  $ indicates that the
mixture is either stable or unstable, and the transition from the stable to
unstable mixture takes place when $\lambda$ exceeds the threshold value
$1+2\varepsilon_{dd}$ irrespective of the particle number densities.
\begin{figure}
[ptb]
\begin{center}
\includegraphics[
width=3.4566in
]%
{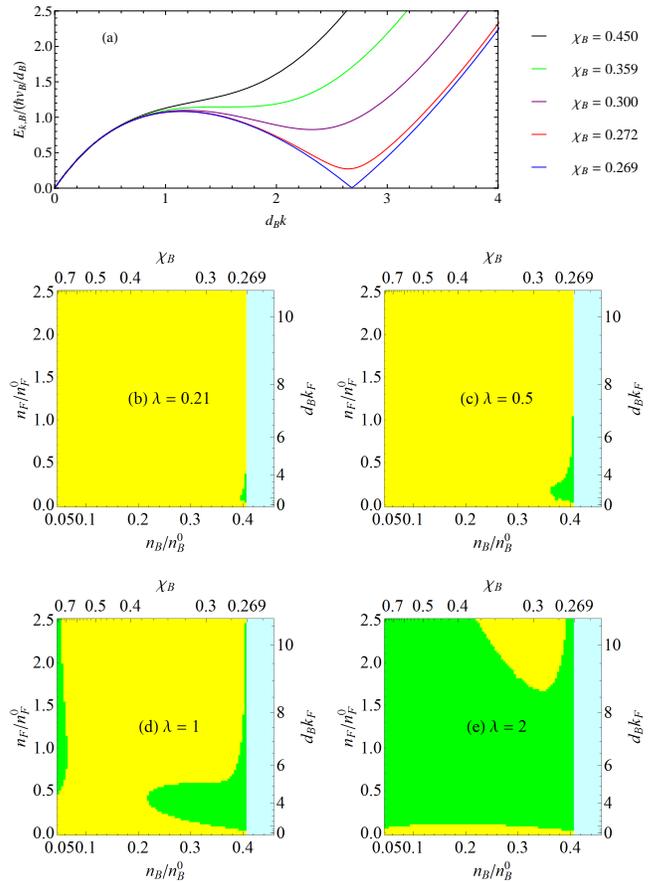}%
\caption{For all diagrams in this article, $\varepsilon_{dd}=2.2$, $m_{F}=6u$,
$m_{B}=127u$, $d_{F}=0.15\mu m$, $d_{B}=0.9\mu m$, and $a_{BB}=100a_{0}$,
where $u$ is the atomic mass and $a_{0}$ is the Bohr radius. \ In addition, a
characteristic Fermi wavenumber $k_{F}^{0}$ is introduced as a basic unit, and
$n_{F}^{0}=\left(  k_{F}^{0}\right)  ^{2}$ and $n_{B}^{0}=2bm_{B}\left(
k_{F}^{0}\right)  ^{2}/m_{F}$\textbf{ }are defined as the scaled number
densities, where $b=\sqrt{\pi\hbar^{2}m_{F}/2m_{B}^{2}U_{BB}}$ is a unitless
parameter. \ Throughout, $k_{F}^{0}$ is fixed to $k_{F}^{0}$\ $=2.11\mu
m^{-1}$ so that $n_{F}^{0}=4.45\times10^{12}$ $m^{-2}$ and $n_{B}%
^{0}=2.99\times10^{14}$ $m^{-2}$($b=1.344$). \ (a) The excitation spectrum,
$E_{\mathbf{k},B}$, is plotted as a function of $d_{B}k$ for $\chi_{B}=0.45$
(black), 0.359 (green), 0.30 (violet), 0.272 (red), 0.269 (blue) (from top to
bottom). (b)-(e) are the stability diagrams in density $\left(  n_{B}\text{,
}n_{F}\right)  $ space for $\lambda=0.21$ (b), 0.5 $\left(  c\right)  $, 1.5
(d), and 2.0 (e) when the Hartree-Fock energies are included. The uniform
quasi-2D mixture is stable (yellow) and unstable (green) against small changes
in densities, and the dipolar condensate is unstable against density waves in
the light blue region where the roton minimum is imaginary.}%
\label{Fig:1}%
\end{center}
\end{figure}

This transition is not a clear cut anymore when the Hartree-Fock energies are
included because $n_{F}$ and $n_{B}$ enter $\bar{U}\left(  k\right)  $ in Eq.
(\ref{phonon-induced interaction}) (via the Fermi wavenumber $k_{F}$ and the
relative healing length $\chi_{B}$) in a complicated manner. Thus, the
functional dependencies of $\mathcal{E}_{exc}\left(  n_{B},n_{F}\right)  $ in
Eq. (\ref{exchange energy density}) on $n_{B}$ and $n_{F}$ can be highly
asymmetrical and nontrivial. \ Figures \ref{Fig:1}(b)-(e) showcase four sets
of stability diagrams in density space with different $\lambda$ when the
Hartree-Fock energies are included. \ As can be\ seen, the region of
instability depends on densities and grows in size as $\lambda$ increases.
\ The density space is mainly occupied by the stable states (yellow) for small
$\lambda$, and by unstable states (green) for sufficiently large $\lambda$.
\ [The light blue area (in which $\chi_{B}$ is lower than $0.269$) is the
region where the roton minimum becomes imaginary.] \ But, the instability,
particularly in the region where the roton minimum is low, develops when
$\lambda$ is far less than $1+2\varepsilon_{dd}$ (which equals $5.4$ for
$\varepsilon_{dd}=2.2$) as indicated by the green areas in Fig. \ref{Fig:1}(b)
and (c).

\section{Chiral $p$-wave superfluid and critical Temperature}

In the parameter space that hosts the single uniform mixture, the interaction
induced by bosons, being attractive, provides identical fermions with the
opportunity to form odd-parity superfluids. \ To investigate this possibility,
we begin with the 2D equation for the gap parameter,
\begin{equation}
\Delta\left(  \mathbf{k}\right)  =-\frac{1}{A}\sum_{\mathbf{k}^{\prime}%
}U\left(  \left\vert \mathbf{k}-\mathbf{k}^{\prime}\right\vert \right)
\frac{\tanh\left(  \beta E_{\mathbf{k}^{\prime}}/2\right)  }{2E_{\mathbf{k}%
^{\prime}}}\Delta\left(  \mathbf{k}^{\prime}\right)  ,\label{gap}%
\end{equation}
which is coupled to the Fock potential, $\Sigma\left(  \mathbf{k}\right)
=-\sum_{\mathbf{k}^{\prime}}U\left(  \left\vert \mathbf{k}-\mathbf{k}^{\prime
}\right\vert \right)  \left\langle \hat{a}_{\mathbf{k}^{\prime}}^{\dag}\hat
{a}_{\mathbf{k}^{\prime}}\right\rangle /A,$ via the quasiparticle energy
spectrum, $E_{\mathbf{k}}=\sqrt{\xi_{\mathbf{k}}^{2}+\Delta\left(
\mathbf{k}\right)  ^{2}}$, where $\xi_{\mathbf{k}}=\epsilon_{\mathbf{k}}%
-\mu_{F}+\Sigma\left(  \mathbf{k}\right)  $ and $\beta=\left(  k_{B}T\right)
^{-1}$. \ In the space spanned by the eigenfunctions of the $z$-component of
angular momentum, $\Delta\left(  \mathbf{k}\right)  =\sum_{l}\Delta_{l}\left(
k\right)  \exp\left(  il\phi_{\mathbf{k}}\right)  $, where $l=0,\pm
1,\pm2,\cdots$ and $\phi_{\mathbf{k}}$ is the polar angle of vector
$\mathbf{k}$. \ In this space, by virtue of the circular symmetry of the
interaction, the gap parameters of different $l$-waves, $\Delta_{l}\left(
k\right)  $, are all decoupled, satisfying%
\begin{equation}
\Delta_{l}\left(  k\right)  =\lambda\int_{0}^{\Lambda}d\epsilon_{k^{\prime}%
}\bar{U}_{l}\left(  k,k^{\prime}\right)  \frac{\tanh\left(  \beta
\xi_{k^{\prime}}/2\right)  }{2\xi_{k^{\prime}}}\Delta_{l}\left(  k^{\prime
}\right)  ,\label{gap l}%
\end{equation}
where $\bar{U}_{l}\left(  k,k^{\prime}\right)  =\int_{0}^{2\pi}\bar{U}\left(
\left\vert \mathbf{k}-\mathbf{k}^{\prime}\right\vert \right)  e^{il\phi}/2\pi$
with $\phi=\phi_{\mathbf{k}}-\phi_{\mathbf{k}^{\prime}}$ and $\Lambda$ is the
energy cutoff (set to twice the Fermi energy $\epsilon_{F}$
\cite{nishida09AnnPhys.324.897} in our estimate below). \ As such, one can
define independently a critical temperature%
\begin{equation}
T_{l}\approx\left(  T_{F}\equiv\frac{\epsilon_{F}}{k_{B}}\right)  \times
\frac{1}{\gamma_{F}}\exp\left[  -\frac{1}{\lambda\gamma_{F}\bar{U}_{l}\left(
k_{F}\right)  }\right]  ,\label{Tc}%
\end{equation}
for each $l$-wave superfluid, where $\bar{U}_{l}\left(  k_{F}\right)
\equiv\bar{U}_{l}\left(  k_{F},k_{F}\right)  $ or
\begin{equation}
\bar{U}_{l}\left(  k_{F}\right)  =\int_{0}^{2\pi}\bar{U}\left(  2k_{F}%
\left\vert \sin\frac{\phi}{2}\right\vert \right)  e^{il\phi}\frac{d\phi}{2\pi
}.
\end{equation}
In arriving at Eq. (\ref{Tc}), we have assumed that only states near the Fermi
energy contribute to the gap equation; we have used the $T=0$ normal Fermi
gas, where the Fermi surface is well-defined, to estimate both the Fermi
momentum, $k_{F}=\sqrt{4\pi n_{F}}$, and the effective mass, $m_{F}^{\ast
}=\gamma_{F}m_{F}$, on the Fermi surface, due to the Fock exchange potential
$\Sigma\left(  k\right)  $, where%
\begin{equation}
\gamma_{F}=\left(  1+\frac{2m_{F}}{\hbar^{2}}\left.  \frac{d\Sigma\left(
k\right)  }{dk^{2}}\right\vert _{k=k_{F}}\right)  ^{-1}.
\end{equation}
Thus, our predictions remain quantitatively accurate only in the weak coupling
regime, but, nevertheless, are expected to shed light on some qualitative
aspects of superfluid pairings under more general circumstances.

\begin{figure}
[ptb]
\begin{center}
\includegraphics[
height=3.1202in,
width=3.1202in
]%
{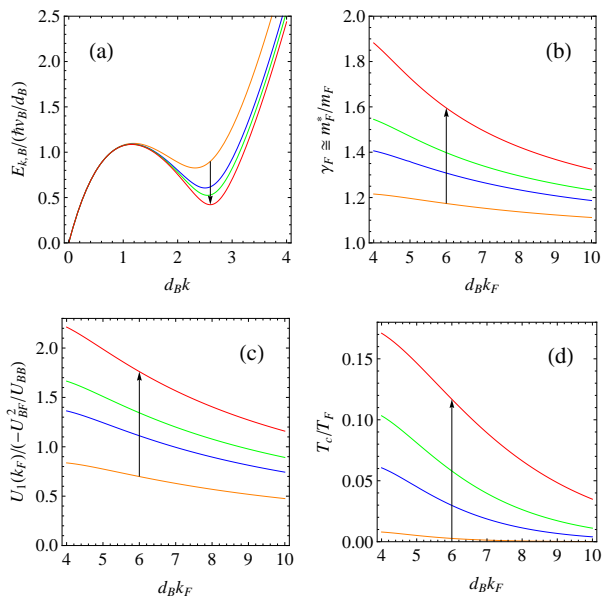}%
\caption{As the roton minimum in (a) is lowered by decreasing $\chi_{B}$ along
the arrow in the sequence of 0.300, 0.284, 0.280, and 0.276, the following
occurs: (b) the ratio of the effective fermion mass to the bare fermion mass,
$\gamma_{F}$, \ (c) the $p$-wave interaction strength, $U_{1}\left(
k_{F}\right)  $, and (d) the critical temperature for the $p$-wave
superfluids, $T_{1}$, all increase as indicated by their respective arrows.}%
\label{Fig:2}%
\end{center}
\end{figure}

In Fig. \ref{Fig:2}, we consider an example in which $a_{BF}=200a_{0}$ which
translates into $\lambda=$ $0.21$. \ A numerical investigation finds that the
$p$-wave pairings dominate other odd superfluid pairings (not shown) when the
product, $d_{B}k_{F}$, exceeds $\sim4$. \ As the roton minimum is lowered
[Fig. \ref{Fig:2}(a)], the $p$-wave interaction, $\bar{U}_{1}\left(
k_{F}\right)  $ (%
$>$%
0), increases [Fig. \ref{Fig:2}(c)], while, at the same time, the effective
Fermi mass also experiences a significant increase [Fig. \ref{Fig:2}(b)].  For a particle in a lattice, its effective mass can be changed by
changing the energy band structure. In our work here, fermions are submerged
in a background of dipolar condensate. \ The (effective) mass of a fermion is
different from its bare value since a fermion is now dressed with phonons of
the dipolar BEC. This polaronic effect, which is typically small and
negligible, is seen to increase appreciably when the roton minimum energy is
lowered [Fig. \ref{Fig:2}(b)].  It
is interesting to point out that while increasing the $p$-wave interaction
surely enhances the exponential factor in Eq. (\ref{Tc}), increasing the
effective Fermi mass also serves to enhance the exponential factor as it
amounts to amplifying the coupling constant, $\lambda,$ by a factor
$\gamma_{F}$ according to Eq. (\ref{Tc}). \ \ These translate into a dramatic
increase in the corresponding critical temperature for the $p$%
-wave\ superfluid phase as clearly demonstrated in Fig. \ref{Fig:2}(d).%

In our system, the induced interaction is independent of the sign of $l:$
$\bar{U}_{l}\left(  k_{F}\right)  =\bar{U}_{-l}\left(  k_{F}\right)  $.
\ Thus, the superfluid with $l=1$ is degenerate with the superfluid with
$l=-1$. A question arises: what is the nature of the $p$-wave superfluid in
our model at $T=0$? \ In principle, the gap parameter becomes the
superposition of these two superfluids, $\Delta\left(  \mathbf{k}_{F}\right)
=\left\vert \Delta_{F}\right\vert \left[  f\left(  \phi_{\mathbf{k}}\right)
\equiv c_{+}\exp\left(  i\phi_{\mathbf{k}}\right)  +c_{-}\exp\left(
-i\phi_{\mathbf{k}}\right)  \right]  $, where $\left\vert c_{+}\right\vert
^{2}+\left\vert c_{-}\right\vert ^{2}=1$. \ The ground state energy due to the
superfluid pairing is then minimized when $\left\vert \Delta_{F}\right\vert $
is maximum \cite{anderson61PhysRev.123.1911}. \ A variational analysis of
$\left\vert \Delta_{F}\right\vert $ obtained from the gap equation (\ref{gap})
at $T=0$ \cite{nishida09AnnPhys.324.897} indicates that only when $c_{+}=0$ or
$1$ does the gap parameter reach its maximum value of $\left\vert \Delta
_{F}\right\vert =2k_{B}T_{1}$ so that $\Delta\left(  \mathbf{k}_{F}\right)
\approx e^{\pm i\phi_{\mathbf{k}}}2k_{B}T_{1}$ (up to a global gauge
transformation). \ Thus, our system realizes at zero temperature the ground
state phase of chiral $p_{x}\pm ip_{y}$ waves whose gap parameters can be
resonantly enhanced by tuning the roton minimum.\ \ 

\section{Conclusion}

In conclusion, we have studied, within the Hartree-Fock-Bogoliubov approach,
the $p$-wave superfluid pairings in a quasi-2D dipolar Bose-Fermi mixture.
\ In a quasi-2D trap setting, the competition between the attractive and
repulsive part of the dipole-dipole interaction can lead to the roton
structure in the phonon spectrum of the dipolar condensate. Mixing dipolar
bosons with fermions in quasi-2D space offers the opportunity to use the roton
minimum as a tool for engineering the phonon-induced attractive interaction
between fermions. \ The present work demonstrates that\ enhancing the induced
interaction by lowering the roton minimum can affect the stability properties
of the mixture as well as the effective mass of the fermions in an important
way. It also shows that one can tune the system to operate in stable regions
where chiral p-wave superfluid pairings can be resonantly enhanced by lowing
the energy cost of the phonons near the roton minimum.

\section*{Acknowledgment}

H.~Y.~L. was supported in part by the US Army Research Office under grant  W911NF-10-1-0096 and in part by the National Science Foundation under grant PHY11-25915.


\end{document}